# Conceptual Framework for Internet of Things' Virtualization via OpenFlow in Context-aware Networks

Theo Kanter, Rahim Rahmani, and Arif Mahmud
*Department of Computer and Systems Sciences, Stockholm University, Sweden*
{kanter, rahim, arifm}@dsv.su.se

## Abstract

A novel conceptual framework is presented in this paper with an aim to standardize and virtualize Internet of Things' (IoT) infrastructure through deploying OpenFlow technology. The framework can receive e-services based on context information leaving the current infrastructure unchanged. This framework allows the active collaboration of heterogeneous devices and protocols. Moreover it is capable to model placement of physical objects, manage the system and to collect information for services deployed on an IoT infrastructure. Our proposed IoT virtualization is applicable to a random topology scenario which makes it possible to 1) share flow-sensors' resources, 2) establish multi-operational sensor networks, and 3) extend reachability within the framework without establishing any further physical networks. Flow-sensors achieve better results comparable to the typical-sensors with respect to packet generation, reachability, simulation time, throughput, energy consumption point of view. Even better results are possible through utilizing multicast groups in large scale networks.

## 1. Introduction

The Internet of Things (IoT) can be outlined in a universal network frame supported by regular and interoperable network protocols in which sensible and virtual "things" are incorporated into the communication network. 'Things', by definition, resembles to any physical object that is capable to interconnect with each other and participate to develop the concept of e-services out of context information received from Internet of Things [1]; The concept of IoT enormously strengthens the service space attainable from the Internet. Establishment of a complete IoT framework can lead to ambient computing and pervasive intelligence through networking and sharing of resources among lots of physical entities in configurable and dynamic networks [2]. A combined cooperation of Internet of Things and OpenFlow is able to hold the dream to attain Infrastructure as a Service and the utmost exploitation of cloud computing.

Availability of context information in modern information systems turns our day to day life simpler and easier. All the devices surrounded us, from any home appliances to any luxuries devices can become responsive of our existence, and mood and can act accordingly [3]. Deployment of flow-sensors in IoT infrastructure can receive context data out of raw data from environment and can lead to play role in development in pervasive computing in such ways

- Information of dynamic environment can be reachable through the placement of static devices.
- Create a better monitoring infrastructure for the systems and services required
- Possibility of dynamic configuration and analysis of the context information and sources.
- Maximum utilization of Internet of Things in terms of reusability, resource sharing and savings.

Present Infrastructure As a Service (IaaS) contain a preset architecture with location aware network mapping along with associated physical devices like different servers and storage devices, routers and switches and running routing logics and algorithms. These topologies cannot support the dynamic one where presence of sensors, intelligent devices are virtual and cannot create a runnable common platform for different kind of traffics. OpenFlow programmability and virtualization feature allows two completely new abstract layers namely common platform layer and virtualization layer to be added at the top and bottom of a preset architecture. It also allows the present infrastructure running without any obstacles even after adding new layer function-abilities. So, only OpenFlow can provide a better solution through network programmability and device virtualizations and thus enable the IaaS to provide the service like security applications, system and network applications, system software etc.

As a continuation of our previous works [4, 5, 6] we have proposed the following ideas to implement in IoT based context aware networks for the sake of achieving a common platform and virtualization with IaaS layer through deploying OpenFlow protocol:

- A completely new idea to merge OpenFlow technology with IaaS layer to make sensor data clouds more efficient from information gaining, sensor management, monitor and virtualization point of view.
- Placement of sensor node is very important for proper data transmission and reception, but in a random scenario, it is almost impossible. Many researchers have proposed highly optimized placement and transmission algorithms but these are too complicated to be implemented practically. So, why should not we try OpenFlow supported flow-sensor?
- Typical Sensor networks are formed in an ad-hoc mode to perform any specific task. So, a common platform is required and OpenFlow is able to provide that even for experimental traffics.
- Data is required to be shared and passed among different wireless objects like sensors, actuators, PDAs etc. OpenFlow is able to provide a common platform and virtualization layer for all networks and thus allow them to share the resources.
- We have suggested a 4 layer conceptual framework to achieve context supported dynamic e-services out of the static Internet of Things. This framework will cover heterogeneous physical entities, placement, integration and synchronization with management system. This context service network will leave the current internet infrastructure unchanged and can be sketched along with diverse systems and services.
- Presently transport layer is only responsible to provide reliability which designates the internet layer to be unreliable and let alone the below layers. But OpenFlow supported sensor are found to be the best candidate since it is delimited to low overhead and multicast assistance as comforted by CoAP application. Besides it can turn the MAC layer more reliable in comparison to typical sensors and so does the network layer.
- At present sensor applications are typically data centric but not the node centric which means we are little concerned about the result of any specific node but the result from the group of sensors.
- Now a day calculation of the number of nodes within the transmission range and packets received are convenient from the stationary nodes viewpoint. But we also need to address the packets from different domain of stationary sensors received by access points.

Possible application includes e-health, home automation, transportation, battle field inspection, safety, failure management and in some other areas where usual and normal attempts were proven to be very expensive and uncertain [7]. Unstructured randomly sited sensors integrated into IoT also have the capabilities to offer a large amount of environmental services such as sound, pressure, temperature, motion etc.

The paper is organized in the following way: Section 2 describes the Motivation and background; Section 3 presents Design and implementation of the proposed model; Section 4 describes the model checking of the new concept; Section 5 presents the performance evaluation and the conclusions are provided in section 6.

## 2. Background

Next generation internet is highly dependable on the incorporation of regular objects found in our surroundings those can be uniquely recognizable, controllable and monitor-able such as sensors, actuators etc. into Internet of Things. Just IP connectivity won't allow wireless sensor network (WSN) to be included in Internet due to their limited resources like bandwidth, memory, energy and communication capabilities [8]. Dynamic internet connectivity happened to be possible through Integration of WSN and IoT. Task evaluation allows gaining benefits from network heterogeneity, remotely accessing becomes possible through efficient collaboration to achieve a certain set of future challenges such as gaining context information from surroundings [9].

Context is a term utilized to distinguish and describe the situation and state of any entity found in our surroundings. It is the information which is considered significant for the communication between users and applications where identity, location, state etc. of the objects are taken into account [10]. Context networks itself can behave as a service since results are collectively collected and turns fault tolerant and effective adaptive system; distribution of service is maintained in the dynamic environment. Inactivity of a few entities doesn't affect largely for the infrastructure and services can be still accessed [11]. Context awareness bears a large prospect in generation of novel services, resource sharing and service quality development in IaaS and dynamic services can be automatically adapted according to context data through changing the service behavior.

Infrastructure as a Service (IaaS) is known to be one of the most important methodologies to communicate with the services offered by cloud computing which preserve applications, information in virtual storages and servers and can be access via web browser from internet [12]. IaaS also provides a solid base to Software as a Service and Platform as a Service. It is responsible to create an abstract layer (vir-tual middleware environment) on physical devices like storages, servers etc. along with offered services. Opportunity is given to user to operate and configure guest OS where storage, bandwidth and other performances matrixes are previously fixed [13, 14].

Network as a service (NaaS) can be an integral part of IaaS through the inclusion of contest aware information where networking loads will be shared, applications will be virtualized and thereby quality of services will be maintained [15]. Up-growing demands of services can be solved through the flexibility and scalability of context supported NaaS. Current communication protocols are maintained and managed by vendor and that's why it is challenging to establish new network services. These network features shouldn't be merged with running protocol so that they can be introduced without changing the current infrastructure. The current network should be adapted to dynamic changes of these services. Network as a service ensures the quality of information and on-demand service through the dynamic configuration of network devices and management [16]. The joint collaboration of network as service and Open-Flow can play a lead role in network virtualization and can maximize the network utilization through resource saving, sharing and distributing among other available nodes or entities.

The OpenFlow can split the traffic path into data packet (maintained by underlying router or switch) and control packet (maintained by a controller or control server) which turn the physical device into a simple one from a complicated mode since complex intelligence programs are removed. Today OpenFlow is supported by several major switch/router vendors (especially a set of functions which are common) and can support all sort of layers (2, 3, and 4) headers [17]. It is also able to integrate the circuit and packet switching technology and these can be treated separately too. Core network also gain noteworthy benefits due to control, management schemes from cost, energy effectiveness and overall network performances point of view [18, 19].

Flow-sensor is just like a typical sensor associated with a control interface (software layer) and flow tables (hardware layer). A Flow table contains a rule (Header) with source and destination address, action that takes the decision (either to drop or to forward the packets) and a counter that maintains a statistics of control and data packet. Control interface exchanges secure messages (control packet) via OpenFlow and sensor buffer maintain typical TCP/IP with access point (data packet exchange) [4].

The ultimate aim of IoT industry is to virtualize and set up a common platform for pervasive computing where context information will be shared and distributed among huge amount of physical entities and create collaboration among multiple services without any centralize system. Within a very short time the Internet of Things industry will be fully established and prepared for large scale manufacture maintaining the services requested by the clients and managing the dynamic changes of the surroundings.

A standard conceptual framework has been proposed taking that in to mind where it can generate context information out of raw data captured from surroundings. This layering concept will also permit new technologies, protocols and services to be introduced and upgrade the IoT technology based on the present infrastructure.

## 3. Conceptual framework

The main tasks of this framework are to analyze and determine the smart activities of these intelligent devices through maintaining a dynamic interconnecttion among those devices. The proposed framework will help to standardize IoT infrastructure so that it can receive e-services based on context information leaving the current infrastructure unchanged. The active collaboration of these heterogeneous devices and protocols can lead to future ambient computing where the maximum utilization of cloud computing will be ensured. This model is capable of logical division of physical devices placement, creation of virtual links among different domains, networks and collaborate among multiple application without any central coordination system.

IaaS can afford standard functionalities to accommodate and provides access to cloud infrastructure. The service is generally offered by modern data centers maintained by giant companies and organization. It is categorized as virtualization of resources which permits a user to install and run application over virtualization layer and allows the system to be distributed, configurable and scalable.

We plan to split the total infrastructure system into 4 layers to receive context supported e-services out of raw data from the Internet of Things. These 4 layers establish a generic framework that does not alter the current network infrastructure but create an interfacing among services and entities through network virtualization. See figure 1.

*3.1. Connectivity layer*

This layer includes all the physical devices involved in the framework and the interconnection among them. Future internet largely depends on the unification of these common objects found everywhere near us and these should be distinctly identifiable and controllable.

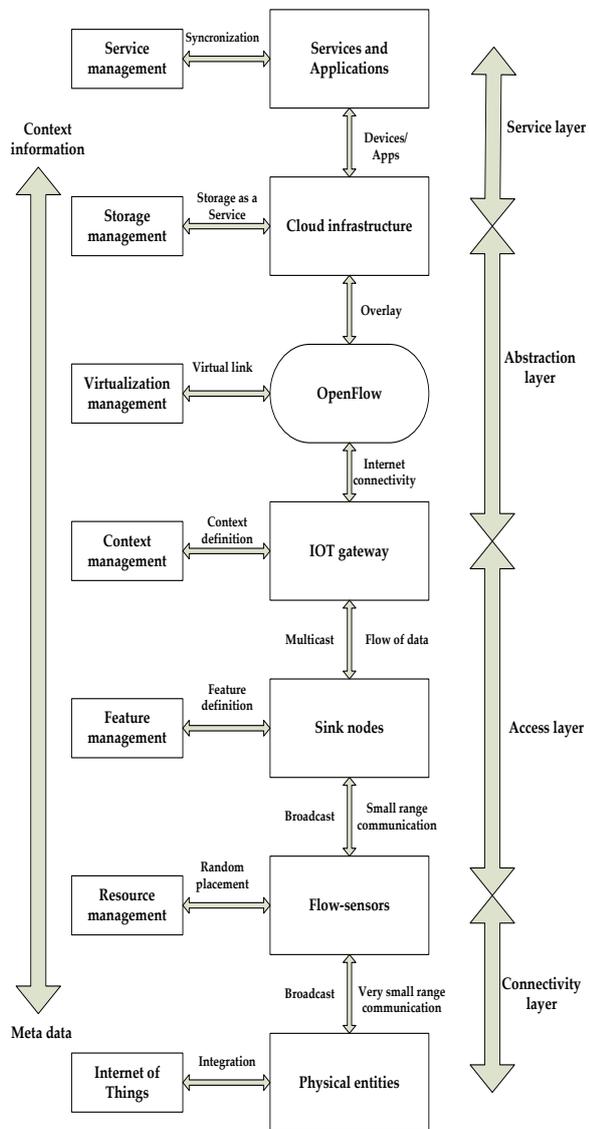

Fig: 1. 4 layer context aware conceptual framework

of raw data. Access layer comprises topology definition, network initiation, creation of domains etc. This layer also includes connection setup, intra-inter domain communication, scheduling, packet transmissions between flow-sensors and IoT gateway. The simulation was run later in this paper for different scenario based on this layer. Feature management contains a feature_filter which accept only acceptable context data and redundant data are rejected. Large number of sensor maintains lots of features but only a small subset of features is useful generate a context data.

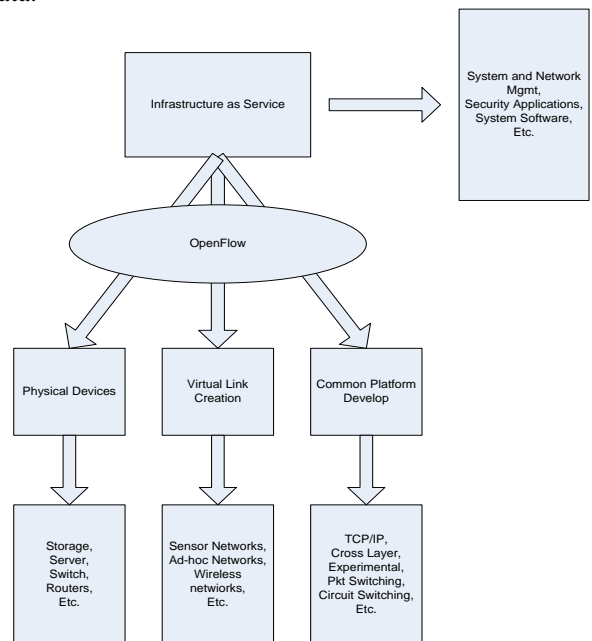

Fig: 2. Infrastructure and services offered

Feature filter helps to reduce irrelevant data transmission, increases the data transfer rate of useful data and reduce energy and CPU consumption too. Number of features can be different based on the application requirements and context data types.

The context management maintains a database which store data received from sink nodes and db controller to check and compare data_values and thres_values and generates action_values. Initially some predefined values are allocated (also known as threshold values), later replaced by newly received values are included (data_values). The database stores only the change value where duplication is not allowed. Database always compares the newly received values with threshold values and creates a decision (action_value) and notifies the IoT Gateway.

*3.3. Abstraction layer*

One of the most important characteristics of OpenFlow is to add virtual layers with the preset layers, leaving the established infrastructure unchanged. As shown in fig. 2, a virtual link can be created among different networks and a common platform can be developed for various communication systems. The system is fully a centralized system from physical layer viewpoint but a distribution of service (flow visor could be utilized) could be maintained. One

This layer also involves assigning of low range networking devices like sensors, actuators, RFID tags etc and resource management checks the availability of physical resources of all the devices and networks involved in the underlying infrastructure. These devices contain very limited resources and resource management ensures the maximum utilization with little overhead. It also allows sharing and distribution of information among multiple networks or single network divided into multiple domains.

RFID tags can be taken example of very short range communicating devices and small enough to be fitted anywhere. It can receive energy from reading object which solves the requirement of battery or external power supply. A large number of RFID tags synchronizes with short range intelligent devices like flow sensors to pass data in a multi-hopping scenario with an aim to reach IoT gateway.

*3.2. Access layer*

Context Data will be reached to internet via IoT Gateway as captured by short range devices in form central system can monitor, control all sorts of traffics. It can help to achieve better band-width, reliability, robust routing, etc. which will lead to a better Quality of Services (QoS).

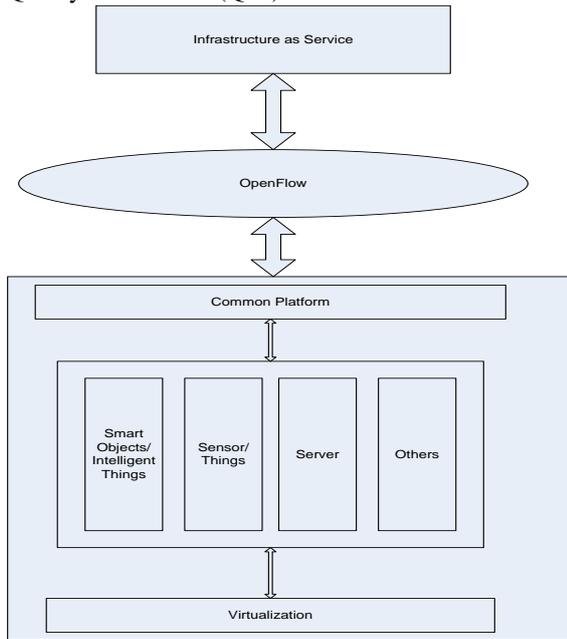

Fig: 3. Three abstraction layers

In a multi-hopping scenario packets are transferred via some adjacent nodes. So, nodes near to access points bears too much load in comparison to distant nodes in a downstream scenario and inactivity of these important nodes may cause the network to be collapsed. Virtual presence of sensor nodes can solve the problem where we can create a virtual link between two sensor networks through access point negotiation. So, we can design a three a three layer platform (fig. 3) where common platform and virtualization layer are newly added with established infrastructure. Sensors need not to be worried about reach-ability or their placement even in harsh areas. Packet could be sent to any nodes even if it is sited on different networks.

*3.4. Service layer*

Storage management bears the idea about all sorts of unfamiliar and/or important technologies and information which can turn the system scalable and efficient. It is not only responsible for storing data but also to provide security along with it. It also allows accessing data effectively; integrating data to enhance service intelligence, analysis based on the services required and most importantly increases the storage efficiency.

Storage and management layer involves data storage & system supervision, software services and business management & operations. Though they are included in one layer, the business support sys-tem resides slightly above of cloud computing service whereas Open-Flow is placed below of it as presented in figure 4 to include virtualizations and monitor management.

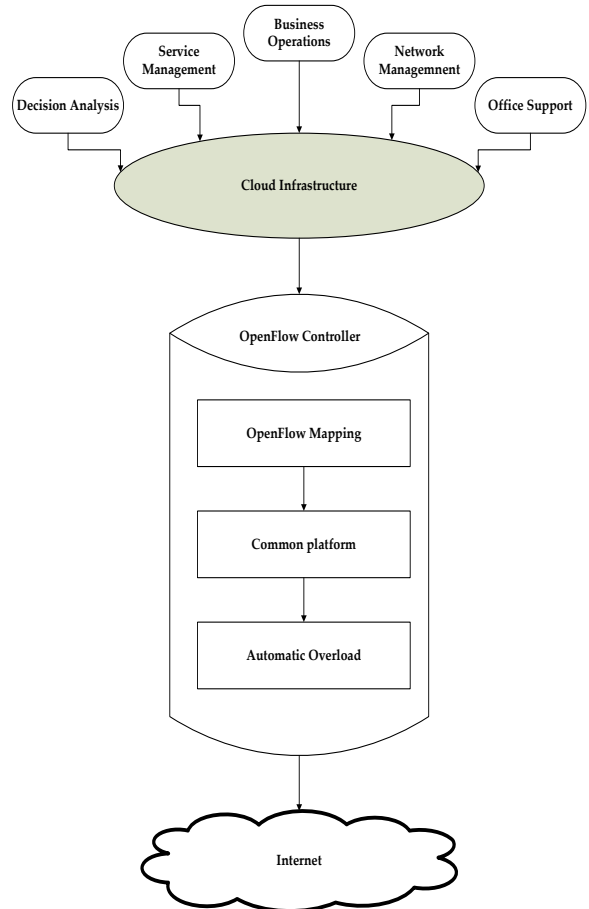

Fig: 4. Cloud infrastructure over internet

All types of business models can find benefits from cloud computing infrastructure. As for example cost and flexibility from small business viewpoint whereas total IT problems can be solved for large companies. It will add advantages for companies, their employees, consumers, distributor where the overall business solution can be provided.

Service management combines the required services with organizational solutions and thus new generation user service becomes simplified. These forthcoming services are necessitated to be co interrelated and combined in order to meet the demand socio- economic factors such as environment analysis, safety measurement, climate management, agriculture modernization etc. [21].

As specified previously any kind of context aware services can be diagramed through this simplified conceptual model. Besides due to heterogeneity management, the framework keeps provision for any technology to be introduced. Most importantly it will leave the established internet infrastructure and cloud technology unchanged and running as well.

**4. Work flow management diagram**

OpenFlow architecture allows building a common platform as found in fig. 5 for different routed-switched traffic and requires to be mapped before that. OpenFlow supports 2nd layer, 3rd layer and even cross layer traffic where source and destination addresses are needed to previously set up. OpenFlow mapping layer establish a connection between physical devices and OpenFlow table via a secured OpenFlow communication protocol. OpenFlow control server generates a tree structure and locates the position of sensor devices. It also can monitor the packet flow in downstream direction and observe the current status of each sensor on a requirement or periodic basis.

## 5. Model checking and concept

The PROMELA and SPIN combination has proved to be a versatile and useful tool in the simulation and verification of software systems [22, 23, and 24]. Both have been extensively used in modeling and verifying communication protocols. In particular, SPIN shall be applied to simulate exhaustively the correctness of flow sensor and provide verification in Linear Temporal Logic (LTL) with respect to convergence.

*5.1. Definition states*

3 different states, shown in fig. 4 Match_pkt, Translation and Send_pkt can be represented by $\Pi$, $\mu$, $\Gamma$ respectively.

Definition 1: Upon receiving, data packet will be matched based on control server state, packet source and packet information. Then the packet will be either dropped or sent to the translation or mapping state accordingly.

$\Pi \models \neg D \square$ Pkt if either $\Pi \not\models \neg D$ or $\pi \models$ Pkt, where receiving of packet = Pkt and packet dropping = D

Definition 2: Translation state maps the data into the flow table and allocates the task into different sensor networks.

$\mu \models F1 \vee F2 \cdots \vee Fn$ for id = {1, 2,⋯, n} and we also achieve $\mu \models Fid$

Different task allocations can be denoted by F1, F2... Fn respectively along with network id as id.

Definition 3: Packet will be sent either to cache or to translation state in case of acknowledgement or data respectively.

$\Gamma \models$ cache $\vee \mu$ iff $\Gamma \models$ cache or $\Gamma \models \mu$, where Add_cache and translation state are represented by cache and $\mu$ respectively.

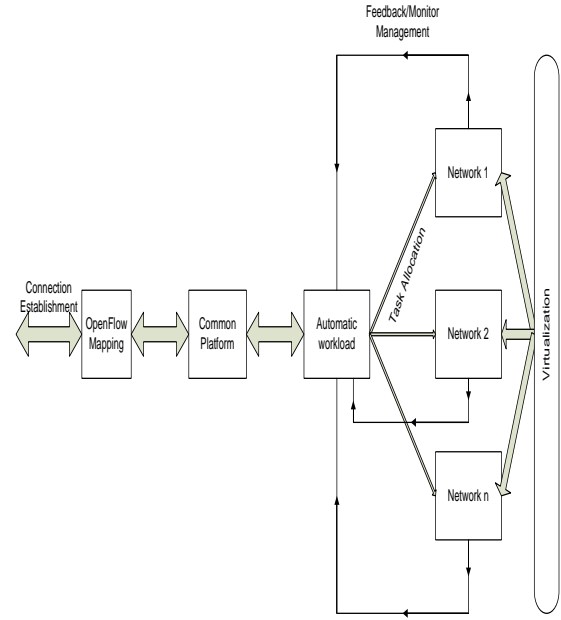

Fig: 5. Work flow management

*5.2. LTL formulas*

The following LTL formulas are generated for the definitions:

LTL1: $\square$ (Receive_pkt $\wedge \neg$ Drop $\square \Diamond$ action ToMatch_pkt)

LTL1: $\square$ (Translation $\square \Diamond$ (action ToTask_alloc 1 $\vee$ action ToTask_alloc 2 $\vee$ …. $\vee$ action ToTask_alloc n))

LTL1: $\square$ (Send_pkt $\wedge \neg$ Drop $\square \Diamond$ (action ToAdd_cache $\vee$ action ToTranslation)

## 6. Packet transmission algorithm

The access node receives the flows of packets from its own and different domain of flow-sensors. The packet transmission algorithm consists of three phases known as network initialization, transmission and reception. Flow table matching and checking have been explained in packet flow algorithm in one of our previous papers [4].

*6.1. Network initialization*

At routing initiation phase, every access point maps all the static nodes connected to it.

pos(x) $\square$ returns the 2D placement of node x.

ap_pos(i) $\square$ returns the 2D placement of access point i.

ap_x(x) $\square$ returns the connected access point of node x.

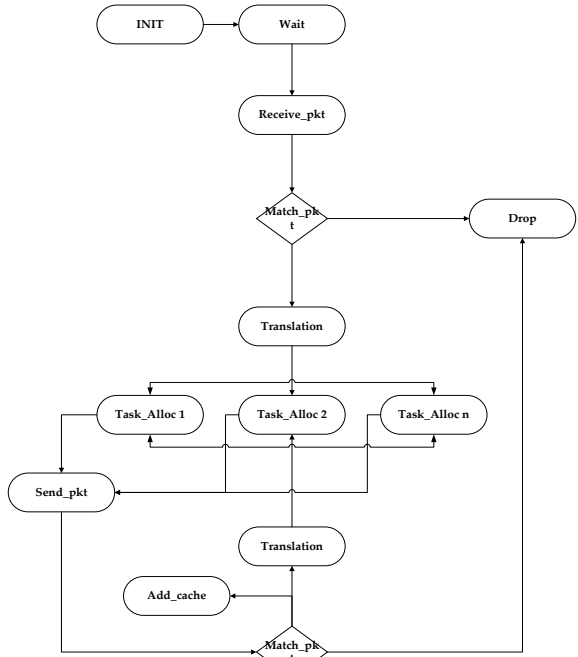

Fig: 6. Data flow diagram

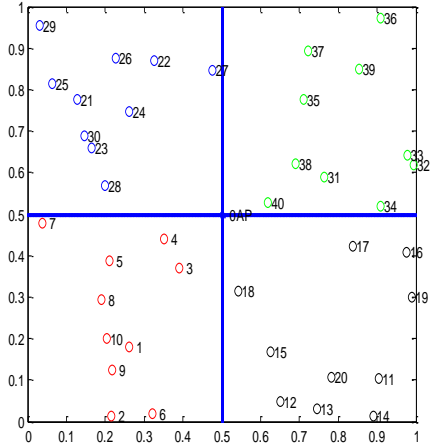

Fig: 7. Four different sensor networks

s_s(x,y) → calculates the sinal strength of transmitting node x to node y.

def_ft(y) → define the flow table of receiving node y.

chk_ft(x) → match the flow table of receiving node y with transmitting node x.

*6.2. Transmission*

Source node, x ε X where X = {source nodes}
    trans_start(x) = schedule(t); // Transmission of transmitting node x starts at scheduled time t.
    for each transmitting node x,
        for each receiving node y,
    trans_range(x); // Calculate the transmission range of node x using Friss model .
    if (distance_required ≥ distance (x,y)) return true,
    X={X,y}; // add the receiving node y with source nodes
    If no new nodes are added to X, return;

    if (y == dest_add); // check if the receiving node is the destination node
        return true;
    trans_end(); // end the transmission
    else return;

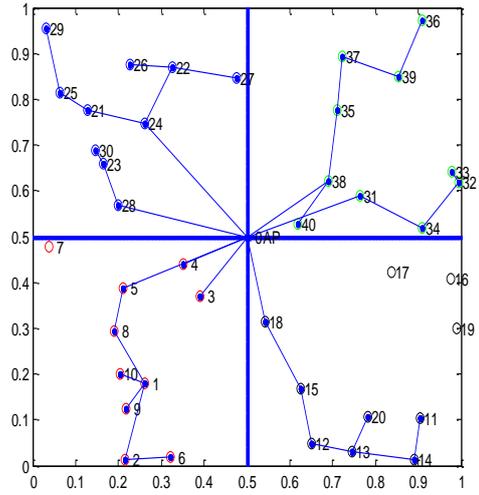

Fig: 8. Individual routing for sensor networks

*6.3. 6.3. Reception*

    schedule (t); // reception starts at time t
    for each receiving node y,
    rcv(bytes); // receive packets from x
    chk_ft (x); // check the flow table of node y with node x
     if( true); trans_start (y); // accept and forward the packet
    else return; // ignore the packet item.

## 7. Reference topology model

The scenario is simulated based on access layer as stated earlier. Total scenario is divided into 2 portions. At first communication is placed between several networks. Then the network got divided into several domains and performance is analyzed based on intra and inter domain communication.

4 different sensor networks have been created with an access point that can be found in fig. 7. All the sensors of different networks will use the access point as the gateway. Different sensor networks are assumed to serve different applications where each network contains 10 sensors. These sensors are randomly sited and access point is situated somehow in the middle. Red, black, green and blue sensors are assumed to the 1st, 2nd, 3rd and 4th network respective-ly.

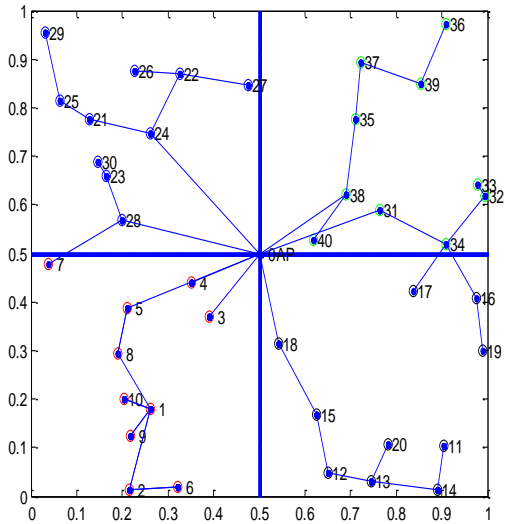

Fig: 9. Sharing resources among different networks

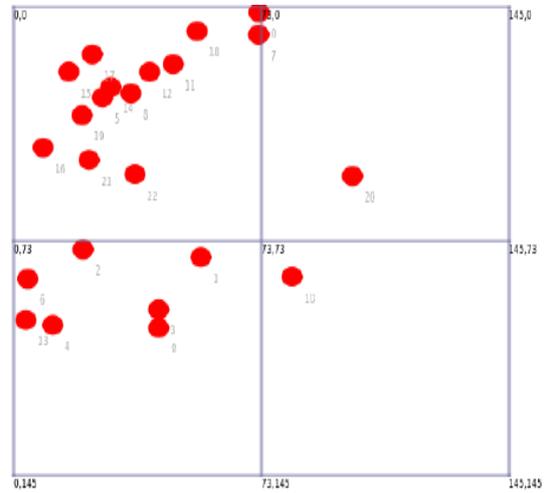

Fig: 10. Random placement of flow-sensors and sink nodes

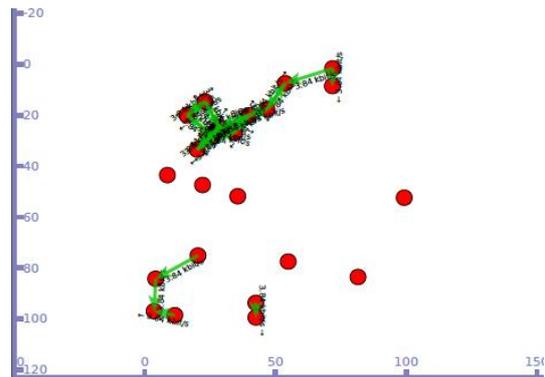

Fig: 11. Creation of multicast domains

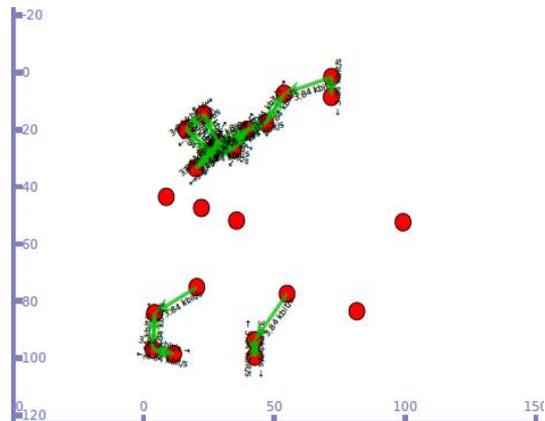

Fig: 12. Communication between sink nodes

At the beginning (fig. 8) all the sensors are assumed to the typical sensors and sensors of one network are not allowed to communicate with other networks. In random networks, some nodes are always sited out of state or away from the range of access point and other sensors. In 1st network node 7 and in 2nd network node 16, 17 and 19 are out of range. That's why we have 100% reach-ability in 3rd and 4th network but 90% and 70% reach-ability in 1st and 2nd network respectively.

Then in fig. 9 all the typical sensors are replaced by flow-sensors and sensors of one network can utilize sensors of other networks for data transfer. We can see node 28 of 4th network can reach node 7 of 1st network. And node 16, 17 and 19 of network 2nd can use node 34 of 3rd network as intermediate nodes in a multi-hopping scenario. So now all the 4 networks are assumed to be a single network virtually and 100% reach-ability can be achieved thereby.

The Ns-3 simulator has been used to simulate the following scenario where IEEE 802.11 was taken as a reference sensor model [25]. IEEE 802.11 is a worldwide accepted model for consumer, public and organizational applications [26]. To be noted Ns-3 is an event supported and going to replace Ns-2 through receiving all the models and features [27]; standard physical and MAC layer functionalities performance has been analyzed in terms of delay, jitter, throughput, energy consumption etc. from various stationary and mobile nodes viewpoint [28], [29] and [30]. We have created a network topology with 24 nodes in fig. 10 where all the nodes are randomly placed.

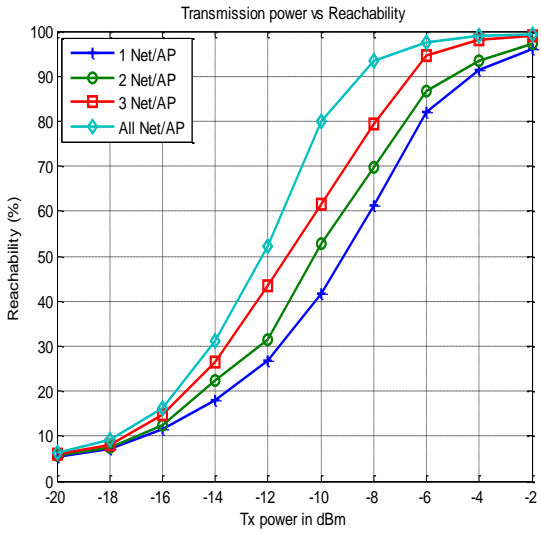

Fig: 13. Reachability comparison on varying transmission power

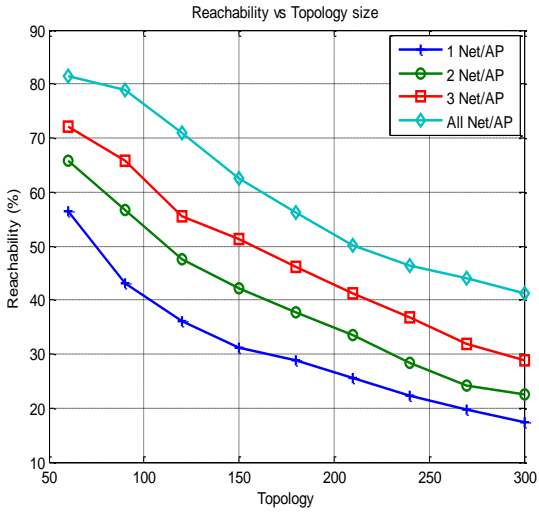

Fig: 14. Reachability comparison on varying topology sizes

4 nodes are acting as sink nodes; node number 0 to 3 where rest them are flow-sensors. Initially sink nodes are not allowed to communication with each other. So, 4 sink nodes have created 4 multicast domains in fig 11. Sink node 0, 1, 2 and 3 contain 11, 0, 3, 1 flow-sensor respectively. Some of the flow-sensors can be in a state of outage and in the scenario. 5 flow-sensors are out of reachability.

Now we want to create a bigger multicast domain where sink node 1 and 3 can communicate with each other and can transfer data as shown in fig. 12. In the same way all sink nodes can be allowed to communicate to create the largest domain. Sink nodes can send data to IoT gateway and it receives data as a set of flows. A single sink node defines the features where IoT gateway generates context data from feature data and sent it to the cloud via internet structure.

## 8. Performance evaluation

The network performance was evaluated based on three different scenarios; inter-network communication, intra-domain communication and inter domain communication.

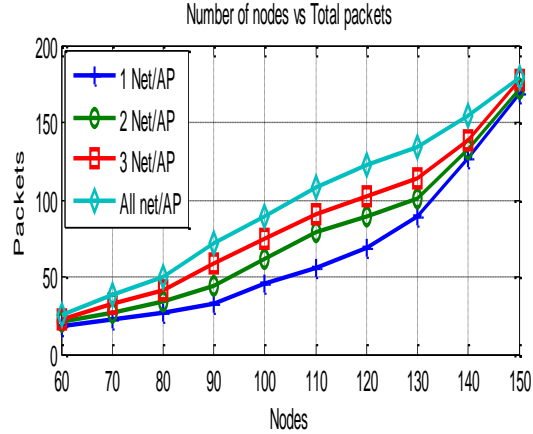

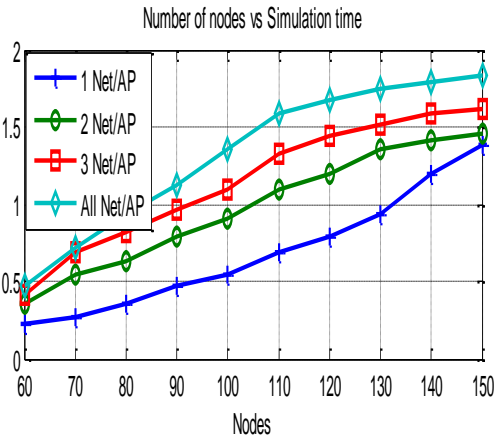

Fig: 15. Total packets and simulation time comparison on varying number of nodes

Fig: 16. Comparison of throughput based on varying node density

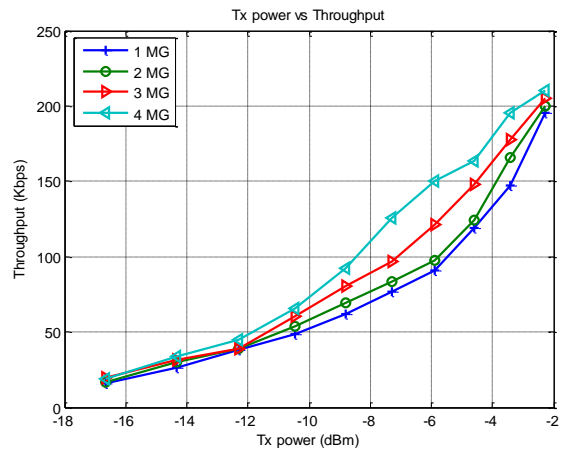

Fig: 17. Throughput evaluation with a changeable transmission power

and inter domain communication. Result analysis was performed following the ideal parameter values by default provided in Table 1 else otherwise noted.

## 8.1. Inter-network communication

The scenario was simulated using Matlab where the metrics include response time and total number of generated packets for varying topology scenario, sensor density and Transmission (Tx) power.

TABLE 1, SIMULATION PARAMETER

| PARAMETER | VALUE OR NAME |
|---|---|
| Communication stacks | RIME |
| Radio model | UDGM & constant loss |
| Node placement | Random 2D position |
| Topology size* | 100*100 |
| Number of nodes* | 100 |
| Sensor density* | 0.01 nodes/meter$^2$ |
| Data rate | 250 Kbit/s |
| Channel check rate | 8 Hz |
| Simulation delay | 0 Sec |
| Maximum retransmission | 15 times |
| Tx range* | 10 meters |
| Interference range* | 10 meters |
| Path gain | -0.04 dBm |
| Propagation constant | 4 |
| Packet size | 125 Byte |
| Required SNR | 4 dB |
| Tx power* | -10.45 dBm |
| Receiver sensitivity | -80.5 dBm |
| Transmission energy | 30 nJ/bit |
| Reception energy | 20 nJ/bit |
| Transmitter antenna gain | 0 dBi |
| Receiver antenna gain | 0 dBi |
| Node mobility | No |
| Simulation runs | 70+ times |

*) will be varied during simulations.

We have compared the performance of flow-sensor and typical sensors where they are randomly sited in maximum four networks with an access point. In 1 Net/AP, sensors of one network are not allowed to communicate with sensors of other networks and they will behave as typical sensors. In 2 Net/AP, 3 Net/AP and 4 Net/AP, sensors of 2 networks, sensors of 3 networks and sensors of all 4 networks will be allowed to communicate as flow-sensors. We have counted the total number of packets (average) and simulation time (total) along with reachability based on varying topology sizes, number of nodes and transmission power.

Fig. 13 explains the reachability of typical sensors and flow-sensors based on varying transmission power. It's true that in a very low and high transmission power both of them behave equally but it is important to know about the ideal scenario activity. All Net/AP reaches 90% of reachability in -8.5 dBm whereas typical sensor requires -4.2 dBm and 2 Net/AP and 3 Net/AP require -5 and -6.8 dBm respectively. In an almost ideal Tx power scenario (-10 dBm), 1 Net/AP, 2 Net/AP, 3 Net/AP and 4 Net/AP have the reach-ability of 41.56%, 52.84%, 61.56% and 79.92% respectively.

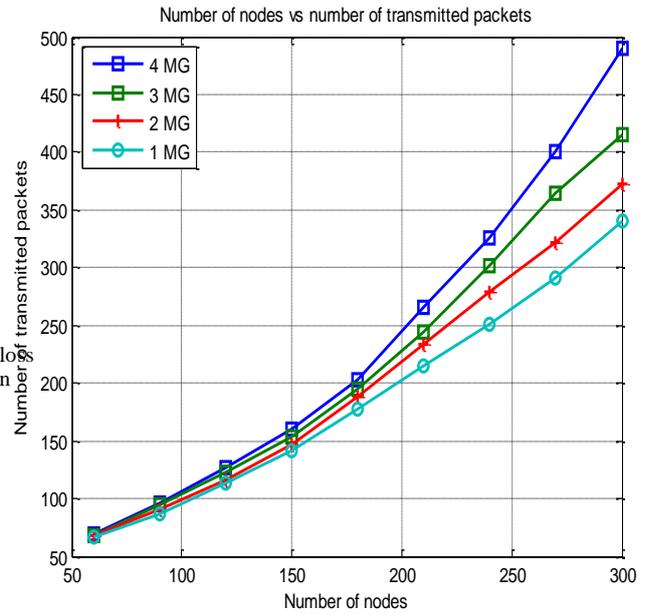

Fig: 18. Comparison of number of transmitted packets based on varying number of nodes

Fig. 14 illustrates reachability counted on varying topology sizes. Reachability of both the typical sensors (1 Net/AP) and flow-sensors (All Net/AP) have been decreased with the increase of topology size. But in all the cases flow-sensors maintains better reachability in comparison to typical sensor network scenario. In a medium scale network (topology size as 180*180), 1 Net/AP, 2 Net/AP, 3 Net/AP and 4 Net/AP have the reach-ability of 28.78%, 37.81%, 46.24% and 56.16% respectively.

In fig. 15 simulation time and total number of packets have been calculated on the same varying amount of nodes. In medium scale networks flow-sensors requires more time to simulate and generate more packets than typical sensors. But in case of higher number of nodes, the difference between them gets decreased. It is true for small networks both of them bear low reachability as reflected in their simulation time and generated number of packets. In a medium scale network (number of nodes = 100), 1 Net/AP, 2 Net/AP, 3 Net/AP and 4 Net/AP have generated 45.23, 61.32, 74.71 and 89.39 packets with a simulation time of 0.55, 0.91, 1.10 and 1.36 sec respectively.

## 8.2. Intra-domain communication

The performance metrics comprises throughput for changing node density and transmission power. UDGM & constant loss has been exploited as a radio model over RIME communication stack to simulate the scenario in Cooja simulator [31]. The problem is addressed by deploying IETF supported IEEE 802.15.4 network model in the physical layer that is capable to operate in low data rate.

The network topology was distributed into 1, 2, 3 and 4 multicast groups denoted as 1, 2, 3 and 4 MG re-spectively. The comparison was carry out based on node density, reachability, transmission power, throughput and maximum number of hops.

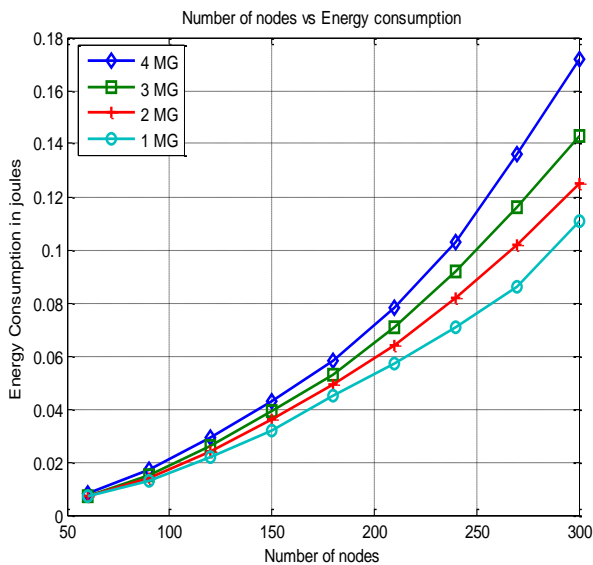

Fig 19: Evaluation of energy consumption with changing number of nodes

As found in figure 16, all of multicast domains had a trivial throughput at low density but initiated mounting up with high density as expected. Small numbers of nodes generate fewer packets and most of packets get dropped due to lower reachability. And it's almost equal for all groups. As a result network efficiency remains lower for all multicast groups at low density. On the other hand packet drop hardly occurs due to high reachability. But packet collision increases with higher number of nodes.

4 MG had a lower packet collision in comparison to other multicast groups that escalated its success rate in highly dense network and so thus the throughput. At node density 0.01 nodes/$m^2$, 4 MG, 3 MG, 2 MG and 1MG accomplished the throughput of 65.32, 60.4, 54.02 and 48.55 Kbps respectively.

Entire groups bear almost equal throughput (very low and very high) at low and high tx power (figure 17).

Reachability has also its effect on the throughput. The throughput increases with the rise and decrease with the decline of reachability. So, it can be claimed that throughput and reachability are proportional to each other in an ideal example (when other factors remain constant).

*8.3. Inter-domain communication*

NS-3 was used to simulate the consequence where the system performance metrics involve total number of generated packets and energy consumption for varying number of nodes.

As seen in figure 18, all of the multicast groups bear almost equal packet generation at the beginning and differences are found with the increase in number of nodes. 4 MG transmits more packets than other multicast groups and seen to be rising for large scale networks.

For 300 nodes, 4 MG, 3 MG, 2 MG and 1 MG transmitted 490.43, 414.72, 372.15 and 340.06 packets respectively.

Fig 19 compares the energy consumption for varying number of nodes. As expected, 4 MG consumes more energy in comparison to other multicast groups for large networks. But the energy requirement differences are very little for the networks of small number of nodes.

4 MG, 3 MG, 2 MG and 1 MG consumed 0.172, 0.143, 0.125 and 0.111J energy respectively in case of 300 nodes.

Reachability can affect both the packet generation rate and energy consumption. To achieve better reachability, more packets are required generated. On the better reachable area, more packets will also be received. As a result more energy will be consumed.

## 9. Conclusion

The proposed context supported framework can systematize IoT infrastructure to receive e-services out of raw data captured by physical devices. The logical division of this model allows to distinct placement of objects, coordination of applications and management functions. A large number of sensors can be divided into groups and send their data to context server which is placed in the clouds via IoT gateway. And the management functions merged with different layers helps to acquire context information from the raw data received from the surrounding.

Context awareness can play a noteworthy role in attaining e-services and pervasive computing as well since it allows interpreting of numerous contexts received from surroundings. The explicit IoT dissection and definite standard allows different manufacturers and system vendors to collaborate their works and large scale development to be fully operational.

Our proposed IoT virtualization can be applicable in a random topology scenario where some of the physical nodes can be sited out of state and inactivity of those nodes can make unreachable from access points. Network virtualization allows flow-sensors of different networks to be used as intermediate nodes under the same platform without establishing any further physical networks. Thus enables resources to be shared, establishment of multi operational sensor networks and escalation of the reachability thereby.

In an inter network communication, All Net/AP achieves more reachability by 18.36, 27.08, 38.36 % points and generate more packets by 14.68, 28.07, 44.16 in comparison to 3, 2, 1 Net/AP in an ideal scenario. On the other hand, 4 MG performed better than other multicast groups in intra and inter-domain communication. 4 MG generate better throughput by 4.92, 11.3, 16.77 Kbps at node density 0.01 node/$m^2$ and more packets by 75.71, 118.28, 150.37 at node density 0.03 node/$m^2$ in case of intra and inter-domain communication respectively. The result trend shows that even better result is possible for large scale networks.

Current network infrastructure cannot handle automatic tuning and adaptive optimization due to the dynamic changes of networks and surroundings. So, utilization of network as a service with OpenFlow technology can bring revolution over present network infrastructure through maximizing the network capacity and fulfilling the demand of dynamic user services and IT solutions specifically from bandwidth, computation power, and storage etc. point of view.

**Acknowledgements**